# Carbon Nanotubes for Interconnect Applications

Franz Kreupl, Andrew P. Graham, Maik Liebau, Georg S. Duesberg, Robert Seidel, Eugen Unger

Infineon Technologies AG, Corporate Research, Otto-Hahn-Ring 6, 81739 Munich, Germany

Tel: +498923444618, email: franz.kreupl@infineon.com

**Abstract**

We briefly review the status of the application of carbon nanotubes (CNTs) for future interconnects and present results concerning possible integration schemes. Growth of single nanotubes at lithographically defined locations (vias) has been achieved which is a prerequisite for the use of CNTs as future interconnects. For the 20 nm node a current density of $5 \cdot 10^8$ A/cm$^2$ and a resistance of 7.8 k$\Omega$ could be achieved for a single multi-walled CNT vertical interconnect.

**Introduction**

As illustrated in Figure 1, CNTs can be thought of being made by rolling up a single atomic layer of graphite to form a seamless cylinder. The resulting structure is called single-walled carbon nanotube (SWCNT). If several SWCNTs with varying diameter are nested concentrically inside one another, the resulting structure is called a multi-walled carbon nanotube (MWCNT). The wave function of the electrons around the circumference is quantized leading to 1-dimensional conductors which can be metallic or semiconducting, depending on how the tubes are rolled up. The semiconducting tubes have a diameter dependent energy gap which roughly scales as 0.9eV/d [nm] [1]. The semiconducting CNTs can be used to fabricate efficient transistors, as shown by our group by prototyping the first nanotube power transistor that can deliver currents in the milliampere regime [2]. Metallic tubes are 1-dimensional metals with a Fermi velocity $v_F$ that equals metals ($v_F \approx 9 \cdot 10^7$ cm/s) and an electron mean free path $l_{fmp}$ for the electrons of at least 1 µm. However, due to the low density of states, the resistivity is only of the order of that of the best metals (~1µΩcm) despite the huge mean free path. The conductance of an ideal ballistic nanotube is $4e^2/h \approx 155$ µS, or about 6.5 k$\Omega$. Additional series resistances needs to be added

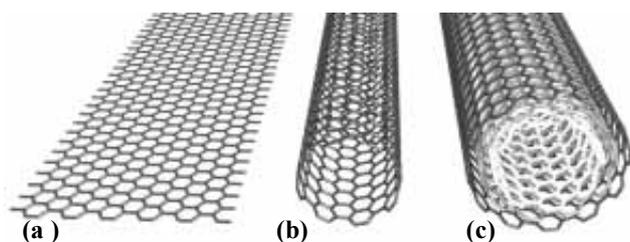

Fig. 1: Carbon nanotubes are made of single graphene sheets. (a) Shows a cut-out part of a graphite lattice. (b) shows a single-walled CNT. (c) Shows a multi-walled CNT, where several CNTs are nested concentrically.

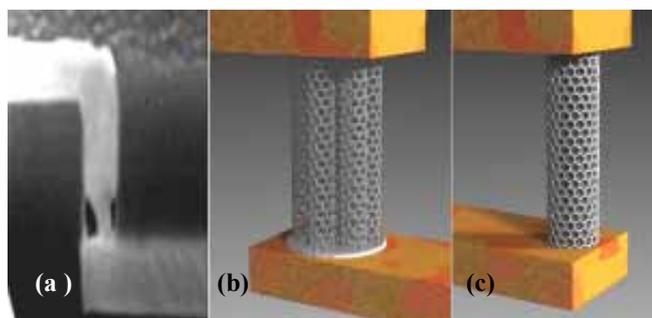

Fig. 2: (a) Conventional Cu-via with voids at the bottom. (b) Proposed replacement of the Cu-via by an array of CNTs. (c) Proposed replacement of the Cu-via by a single MWCNT.

Table I

Reliability data from 2 nanotubes measured at 250°C for 350 hours (adapted from [5]). Data for copper wires are not available for these dimensions, instead the calculated resistances for Cu-wires are shown.

|  | Nanotube I | equivalent Cu-wire I | Nanotube II | equivalent Cu-wire II |
|---|---|---|---|---|
| Diameter | 8.6 nm | 8.6 nm | 15.3 nm | 15.3 nm |
| Length | 2.6 µm | 2.6 µm | 2.5 µm | 2.5 µm |
| Current | 10.4 mA |  | 10 mA |  |
| Resistance | 2.4 k$\Omega$ | 5.6 k$\Omega$ | 1.7 k$\Omega$ | 1.0 k$\Omega$ |
| Current density | $1.8 \cdot 10^{10}$ A/cm$^2$ | ------ | $5.4 \cdot 10^9$ A/cm$^2$ | ------ |

for the contact resistance $R_C$ and the Drude-like length dependent resistance $h/(4e^2)(L/l_{fmp})$ which accounts for additional backscattering for a nanotube with length L. For MWCNTs each individual shell can contribute to the conductance and larger diameters can add additional channels to the conductance [3].

**Resistance and Current density**

For the implementation as future interconnects, CNTs have to fulfill the requirements of high current carrying capability and resistances comparable to or better than copper in vias or tungsten in contact holes. As shown in Table 1, CNTs can withstand current densities up to $10^{10}$ A/cm$^2$, exceeding copper by a factor of 1000. With respect to resistance, CNTs are favorable in high aspect ratio structure like vias, where also the highest current densities are expected. Figure 3 gives a rough estimate for obtainable resistances for CNTs as compared to copper. The resistance of cylindrical via has been calculated by assuming a length independent conductance of $4e^2/h$ for each CNT shell and the size dependent resistivity of copper [4]. Depending on the diameter, a via can be filled with a single MWCNT, a densely packed array of MWCNTs or densely packed arrays of SWCNTs (See Figure 2), which would give the lowest resistance. Additional doping of the CNTs can lower the curves for MWCNTs by at least a factor of 0.5 because higher sub-bands can contribute to the conductance [6].

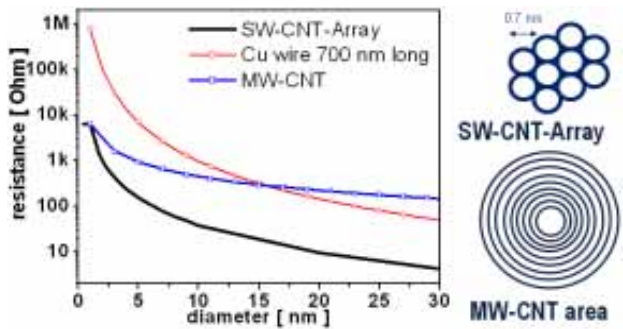

Fig. 3: Theoretical resistance for via fillings with Cu and an array of SWCNTs or one MWCNT with varying diameter. The array of SWCNTs would give the lowest resistance.

### Integration of carbon nanotubes

A catalyst particle which facilitates the growth and determines the diameter of the CNT is usually required to grow CNTs, although there have been some reports of catalyst free formation of CNTs by microwave and template assisted growth methods [7,8]. Filling via structures with nanotubes requires one ore more particles at the bottom of the via which then allow CNT growth by chemical vapor deposition (CVD) at 450-800°C with a carbon containing gas. The CVD process can be supported by plasma enhancement (PECVD) and bias voltage [9]. The catalyst material (Fe, Ni, Co or combinations of them with Mo) is usually deposited as a thin film by physical vapor deposition (PVD) or from solutions. Particle formation occurs during the heating step, which breaks up the thin film into clusters. Ion bombardment in plasmas supports this particle formation. Careful material and interface design in combination with low temperature budget and time dependent diffusion phenomena, needs to be taken into account to guarantee CNT growth [9]. The interaction between catalyst layer and supporting metal electrode needs to be low, in order to suppress interdiffusion and allow particle formation in the restricted temperature regime. Metals with a natural thin oxide layer (Ta, Al, Ti, Cu, Cr) show low wetting behavior for some catalyst materials and are therefore suitable as electrode materials.

### The quality of nanotubes

The quality of the grown CNTs determines the conductivity of tube and depends on the number of shells, their parallelism with the CNT axis (See Figure 4 and Figure 5) and possible defects. Examples for different qualities are shown in Figure 5. CNTs grown by PECVD often show serious deviation from perfect structure especially when they are grown below 650°C [9]. Their resistivity can be described by a model (See Figure 4) that also applies for carbon fibers and is typically in the range of 4 mΩcm [10]. The carriers have to hop from shell to shell which reduces the conductivity tremendously because the conductivity along the c-axis (from shell to shell) is a factor of 1000 smaller than the in-shell conductivity. Ballistic transport can only be expected with well aligned shells. For a given diameter, the conductivity increases with the number of participating shells [3,11]. Therefore, the hollow space within the tube should be reduced to a minimum.

### Integration schemes

Currently two different approaches have been investigated by three groups. Li et al. [12] have proposed a so called bottom-up approach, shown in Figure 6. Here, the CNT via is grown on the metal 1 layer before the deposition of the inter-metal dielectric (IMD). Lithographically defined nickel spots act as catalyst particles, from where carbon fibers are grown. As a prerequisite, the fibers need to be aligned perpendicular to the surface. This is achieved by PECVD and an applied bias voltage, which aligns the fibers almost perpendicular to the wafer. Subsequently, $SiO_2$ is deposited and the wafer is planarized with chemical wafer polishing (CMP). The last step also opens the nanotube ends for contacts with metal 2 layer. A very high resistance of ~ 300 kΩ per CNT interconnect has been evaluated for that approach, which may be attributed to the imperfect structure of PECVD grown MWCNTs. The approach is especially suited for single MWCNT fillings because high density growth could not be demonstrated. In addition, this approach can also be used to create high aspect ratio capacitor electrodes for DRAM applications.

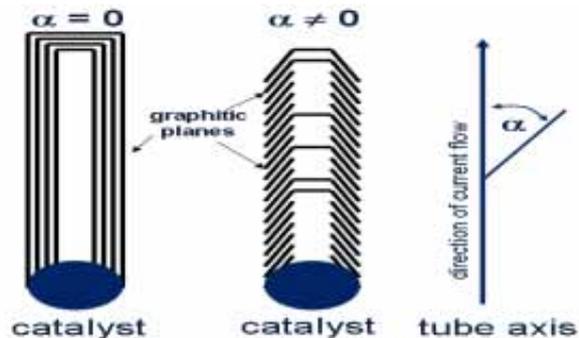

Fig. 4: Factors influencing the resistivity of a tube: number of shells and orientation of the graphitic planes [10]. Ballistic transport can only occur for $\alpha = 0°$, For graphite $\rho_c = 1000 \cdot \rho_a$ and the averaged resistivity can be expressed as $\rho(\alpha) = \rho_a \sin^2(90° - \alpha) + \rho_c \cos^2(90° - \alpha)$ [10].

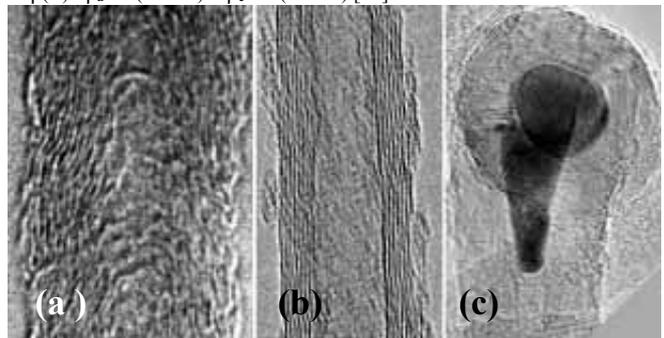

Fig. 5: TEM images for CVD-grown MWCNTs. The interlayer distance is 0.34 nm (a) A poor quality tube grown at 450°C with $\alpha \approx 18°$. (b) A high quality tube with $\alpha \approx 0°$, but with only 7 shells. (c) A high quality tube with ~25 shells also showing the encapsulated catalyst particle.

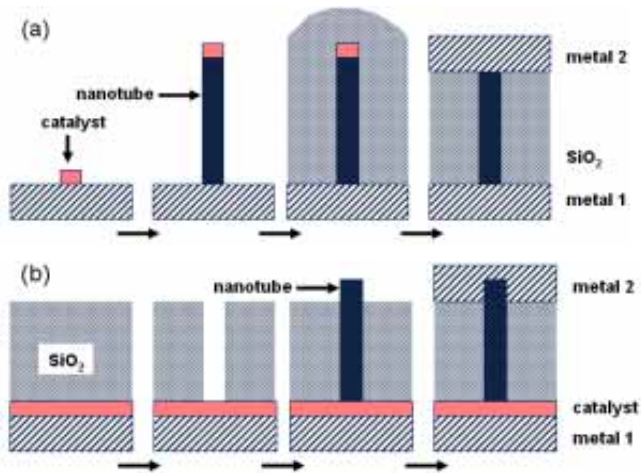

Fig. 6: Possible integration schemes: (a) Bottom-up approach [8].(b) Buried catalyst approach [3,9].

The more conventional approach of etching the vias down to metal 1 layer and growing the CNTs in these vias (See Figure 6(b)) has been undertaken by Nihei et al. and our group. Nihei et al. have used a buried catalyst approach, where the dry-etching of the via has to stop on the thin Ni- or Co-catalyst layer [13]. Arrays of MWCNTs have been grown in ~2μm diameter vias by hot-filament CVD (HF-CVD). A resistance of ~134 kΩ per MWCNT has been achieved, a value which again can be attributed to the quality of the tubes grown by HF-CVD.

In contrast, we have used a pure CVD approach to grow high quality tubes in vias exhibiting resistances of ~10 kΩ per

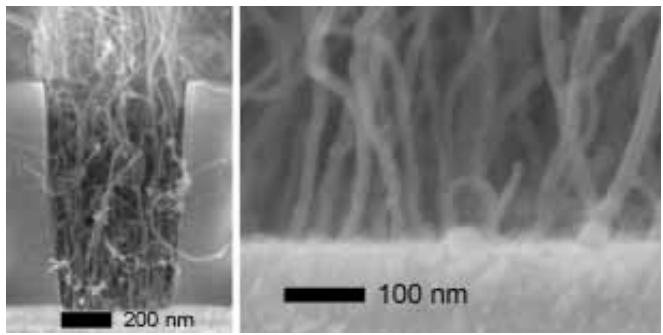

Fig. 7: Cross-section SEM images of vias with CVD-grown MWCNTs. The catalyst particles are based on Fe and the supporting metal layer is Ta.

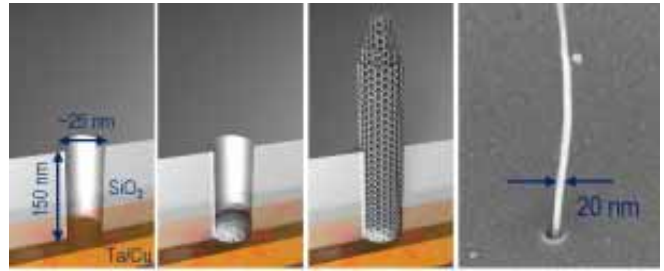

Fig. 8: Fabrication of end-of-the-roadmap sized MWCNT vias with ~20 nm in diameter and 150-200 nm in depth by e-beam lithography and dry-etching using hardmasks. A 20 nm diameter MWCNT is protruding from the via.

MWCNT [3]. In order to manage the via etch stop on a ~2 nm catalyst layer, a catalyst multilayer stack has been developed, which allows proper landing on the catalyst layer with reliable growth of MWCNTs with a density varying between 100-1000/μm$^2$. Figure 9 gives an example of a densely filled 400 nm diameter MWCNT via.

### Nanotube interconnects for the 20 nm node

CNT interconnects for the 20 nm node have been fabricated, using the buried catalyst approach, as shown in Fig. 7(b) [14, 15]. The procedure, dimensions and resulting MWCNT via is shown in Figure 8. A catalyst layer consisting of a triple stack of (3 nm Fe, 5nm Ta) has been deposited on metal 1 prior to the deposition of 150- 200 nm $SiO_2$. Electron beam lithography in combination with a hard mask technique has been used to define the vias. The 20 nm diameter vias have been etched in an MERIE reactor. Subsequently, MWCNTs have been grown by CVD at 450-700°C. The quality and yield of the CNT-vias rises with growth temperature. In the temperature regime below 550°C, only low quality CNTs, such as shown in Fig. 5(a), can be grown. Before deposition of the top contact (metal 2 layer), the MWCNTs have been encapsulated by a ~20 nm thin $SiO_2$ either by spin-on-glass or PECVD deposition.

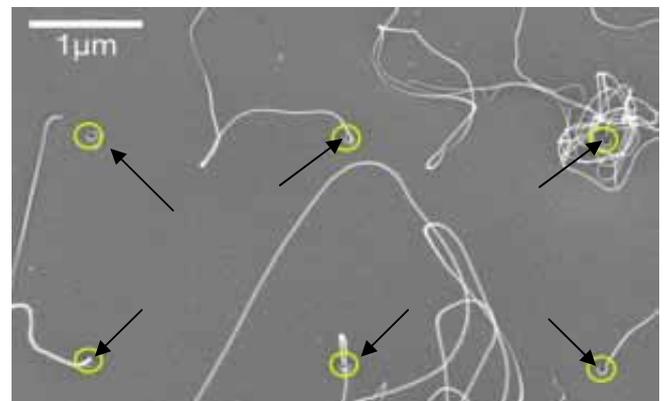

Fig. 9: SEM image demonstrating the yield in the formation of vertical interconnects. MWCNTs protruding from six nano-sized vias in the field of view [14]. The vias have a spacing of 2 μm to allow for easy contacting.

The end of the MWCNT is revealed again by removing excess SiO$_2$ by CMP or a back etch. This step prevents metallic whisker formation during top contact preparation which may affect the resistance measurements. Figure 9 gives an example of the reproducibility of this approach. Various metals such as titanium, palladium and tungsten have been used as a top contact and the achievable resistances examined. The measurements show that an annealing step is required to yield the lowest possible resistance as shown in Figure 10. Here, the I(V)-curve of a MWCNT contacted with tungsten is shown after a final annealing step at 850°C.

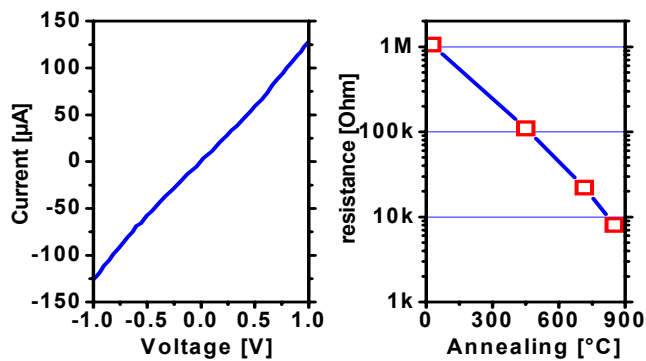

Fig. 10: I(V)-characteristic of a ~20 nm wide MWCNT via with a resistance of 7.8 kOhm and current density of $5 \cdot 10^8$ A/cm$^2$. The annealing dependent resistance demonstrates the difficulty in obtaining good contact resistances.

The drastic change in resistance can be attributed to the improvement of the bottom and top contact by annealing in a reducing atmosphere. A statistical comparison of top contacts made of Pd and Ti in Figure 11 demonstrates better top contacts for Pd. The overall resistance after annealing is still governed by the ability to contact all shells. This again is related to the details of the back etch or CMP processes, which are used to planarize the wafer. Proper contacts can reduce the resistance to about 120 Ω/μm [11].

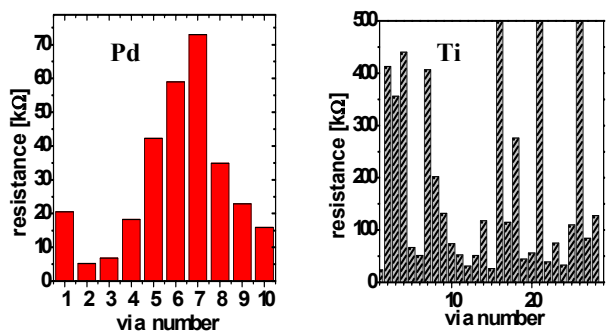

Fig. 11: Statistics for single MWCNT via resistances with 15-25 nm in diameter and 200 nm in height for Pd and Ti contacts after annealing. Pd contacts will give the lowest contact [15].

In summary, the measured current densities of $5 \cdot 10^8$ A/cm$^2$ in these MWCNT vias exceed already the achievable values for metals, whereas the obtained resistance of 7.8 kΩ still has to be lowered by 2 orders of magnitude to compare with theoretical values for copper.

## Acknowledgements

We would like to thank W. Hoenlein for continuous support, W. Pamler for artwork and Z. Gabric for expert technical assistance. The work is supported by BMBF under contract # 13N8402.